\documentclass[%
reprint,superscriptaddress,
amsmath,amssymb,
aps,
]{revtex4-2}
\usepackage{braket}
\usepackage{graphicx}
\usepackage[upgreek, LGRgreek]{mathastext}

\begin{document}


\title{Bright and photostable single-photon emitter in silicon carbide}


\author{Benjamin Lienhard}
\email[]{blienhar@mit.edu}
\affiliation{Department of Electrical Engineering and Computer Science, Massachusetts Institute of Technology, Cambridge, MA 02139, USA}
\affiliation{Department of Information Technology and Electrical Engineering, ETH Z{\"u}rich, Gloriastrasse 35, CH-8092 Z{\"u}rich, Switzerland}

\author{Tim Schr{\"o}der}
\affiliation{Department of Electrical Engineering and Computer Science, Massachusetts Institute of Technology, Cambridge, MA 02139, USA}

\author{Sara Mouradian}
\affiliation{Department of Electrical Engineering and Computer Science, Massachusetts Institute of Technology, Cambridge, MA 02139, USA}

\author{Florian Dolde} 
\affiliation{Department of Electrical Engineering and Computer Science, Massachusetts Institute of Technology, Cambridge, MA 02139, USA}

\author{Toan Trong Tran}

\author{Igor Aharonovich}
\affiliation{School of Mathematical and Physical Sciences, University of Technology Sydney, Ultimo, NSW 2007, Australia}

\author{Dirk Englund}
\affiliation{Department of Electrical Engineering and Computer Science, Massachusetts Institute of Technology, Cambridge, MA 02139, USA}



\date{\today}

\begin{abstract}
	Single-photon sources are of paramount importance in quantum communication, quantum computation, and quantum metrology. In particular, there is great interest in realizing scalable solid-state platforms that can emit triggered photons on demand to achieve scalable nanophotonic networks. We report on a visible-spectrum single-photon emitter in 4H silicon carbide (SiC). The emitter is photostable at room and low temperatures, enabling photon counts per second in excess of 2$\times$10$^6$ from unpatterned bulk SiC. It exists in two orthogonally polarized states, which have parallel absorption and emission dipole orientations. Low-temperature measurements reveal a narrow zero phonon line (linewidth $<0.1~$nm) that accounts for $>30~$\% of the total photoluminescence spectrum.
\end{abstract}


\maketitle

\section{Introduction}
Efficient, on-demand, and robust single-photon emitters (SPEs) are of central importance to many areas of quantum information processing~\cite{2013.AcademicPress.Migdall.SPG}. Color centers in high-bandgap semiconductors have emerged as excellent SPEs~\cite{2010.Wiley.Santori.SPE, 2005.RPP.LounisOrrit.SPE} that can operate even at room temperature. These include atomic defects in diamond~\cite{DDia}, zinc oxide~\cite{DZnO,Neitzke2015_ZnO}, and single rare earth ions~\cite{DRE} in garnets. Color centers in silicon carbide (SiC) and diamond provide optical access to internal spin states~\cite{2015.Springer.Naydenov.SPED,SiVSiC2}, which can be used as quantum memories for quantum computing~\cite{2010.PNAS.Awschalom.QuanDef}, quantum-enhanced sensing, and other quantum technologies~\cite{childress2014atom,doherty2013nitrogen}. 

SiC is a technologically important material that is widely used in optoelectronics, high-power electronics, and microelectromechanical systems~\cite{SiCTod}. It is commercially available in up to 6 in. (152.4$~$mm) wafers, and processes for device engineering are well developed~\cite{SiCProt}. Recently, a large variety of SPEs were reported in SiC and attributed to the carbon antisite-vacancy pair~\cite{SiCRT}, silicon vacancies~\cite{SiVSiC2, SiVSiC4,2015.NatMat.Widmann.SpinSiC, 2015NatCom.Fuchs.SiVSpin}, and divacancies~\cite{DiV1, DiV2, DiVCoh}. 

Here, we report on an exceptionally photostable and bright SPE in the hexagonal 4H-SiC polytype with photoluminescence (PL) in the visible spectrum. Despite the high index of SiC (2.65 at 600$~$nm~\cite{1971.Shaffer.RefIn}) and the associated total internal reflection at the SiC interface, we measure a saturated photon count rate of up to 2$\times$10$^6$$~$counts per second (cps) from individual emitters, representing one of the brightest SPEs~\cite{2011.NatPhot.Loncar.NVEnhanced, SiCRT} from an unpatterned bulk semiconductor. The emitter is exceptionally photostable: we observe no blinking at time scales between 100$~\mu$s to several minutes, and no degradation in the PL properties after months of imaging the same emitter.

\section{Experiments and Results}
\begin{figure}[ht!]
	\centering
	\includegraphics[width=8.5cm]{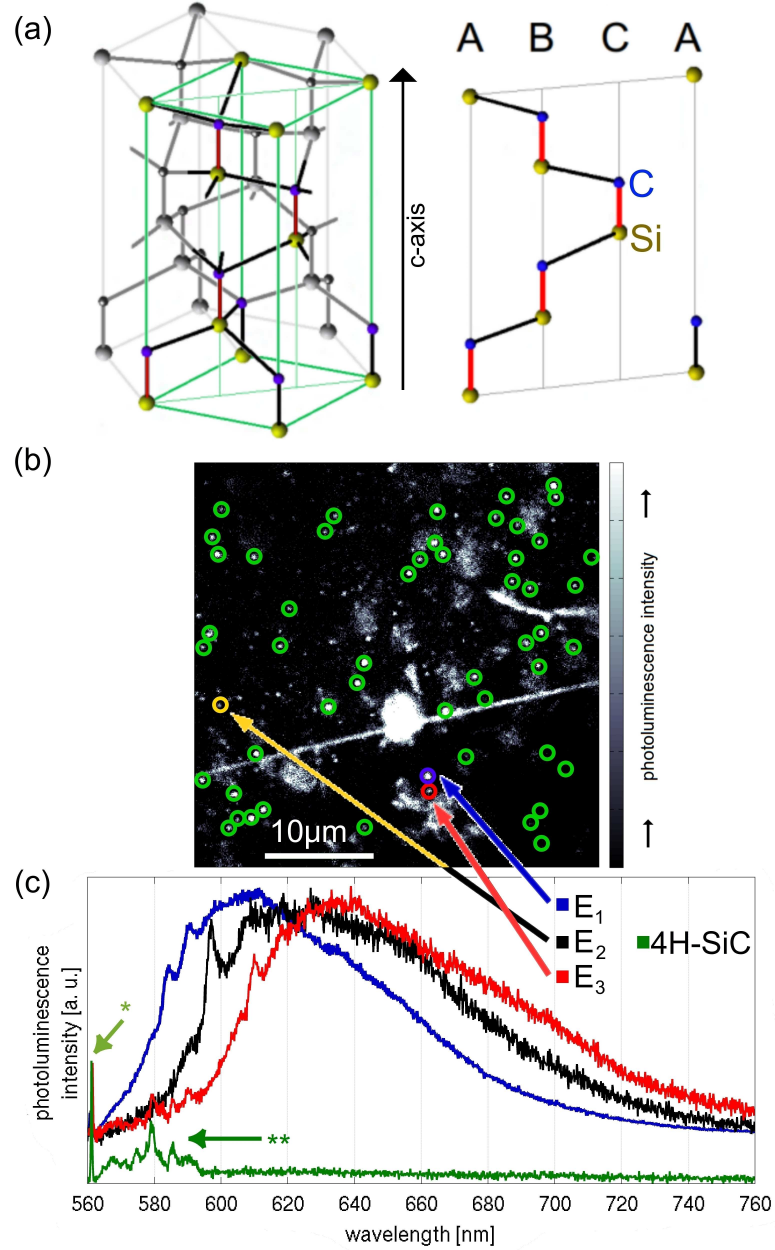}
	\caption{(a) Three-dimensional 4H-SiC crystal lattice [blue, carbon (C); yellow, silicon (Si)] in a hexagonal geometry with one green highlighted rectangular prism containing one two-dimensional layer in its vertical diagonal plane. This plane is shown on the right-hand side to illustrate the one-dimensional lines A, B, and C. The characteristic sequential bilayers in 4H-SiC are ABCB. (b) The confocal fluorescence scan of the top surface of 4H-SiC shows a high density of bright emitters (green circles), which were separately confirmed to have single-photon emission characteristics (see Fig.~\ref{LevelAnti}). (c) Spectra from the three characteristic 4H-SiC emitters (blue, E$_1$; black, E$_2$ [yellow in (b)]; red, E$_3$) indicated in (b), representing the PL distribution over an energy range of 100$~$meV at room temperature. The green spectra, PL of 4H-SiC, shows the characteristic first-order longitudinal optical Raman mode (labeled on the spectrum with a star) and the second-order Raman modes (labeled with two stars).}
	\label{Figure1}
\end{figure}

For our experiments, we use a semi-insulating high-purity (resistivity of more than 100$~$k$\Omega$cm) single-crystal 4H-SiC 3 in. (76.2$~$mm) wafer with a thickness of 350$~\mu$m (W4TRE0R-0200, CREE, Inc). The top surface is the $\{0001\} \pm 0.25^\circ$ crystal plane. Fig.~\ref{Figure1}(a) shows the 4H-SiC lattice structure, which is characterized by sequential layers ABCB, where (A, B, C) denote the SiC bilayer structures, composed of three atoms connected by two bonds (Si-C-Si), that form the basis of any SiC polytype. Layers A and B differ only in a lattice shift, whereas layer C describes a lattice twist by 60$^\circ$. 

The sample was annealed in forming gas (H$_2$:N $\rightarrow$ 1:19) at ambient pressure at 600$^{\circ}$C. Irradiating the sample with electrons (density: $10^{17}$$~$e$^{-}$cm$^{-2}$, energy: 10$~$MeV) before annealing increased the emitter density by $\sim 2\times$.

\subsection{Photoluminescence at Room Temperature}
After the annealing step, scanning confocal microscopy revealed individual color centers emitting in a broad spectrum covering $\sim 580 - 700~$nm, as shown in Figs.~\ref{Figure1}(b) and \ref{Figure1}(c), under laser excitation (532$~$nm, 0.5$~$mW on sample) with a 560$~$nm long-pass filter. The PL was collected with a 0.9$~$NA microscope objective (Nikon) and detected on avalanche photo diodes (APDs) via single-mode fibers. Antibunched photon emission was confirmed by Hanbury Brown-Twiss (HBT) interferometry, as described below. Our measurements showed no evidence of blinking at time scales from 100$~\mu$s to several minutes, and the emission is stable for hours under continuous laser excitation. 

The room-temperature PL spectra extend from $\sim590$ to $700~$nm ($\sim 1.77 - 2.1~$eV), with energy shifts up to $\pm 50~$meV between different emitters. This can be seen from Fig.~\ref{Figure1}(c), which plots the individually normalized PL spectra of the three characteristic emitters E$_1$, E$_2$, and E$_3$, labeled in Fig.~\ref{Figure1}(b). The small PL peaks between $\sim 580$ and $590~$nm are also visible in the background spectrum, shown in the green curve in Fig.~\ref{Figure1}(c), and are ascribed to second-order Raman shifts~\cite{Ram2SiC}, further discussed in Supplement 1. 

We attribute the peaks at 584.8$~$nm (E$_1$), 597.3$~$nm (E$_2$), and 609.6$~$nm (E$_3$) to the room-temperature zero phonon lines (ZPLs) of these emitters, though the assignment is less clear for E$_1$ because of the proximity to the Raman peaks. 

\subsection{Level System}
\begin{figure*}[ht!]
	\centering
	\includegraphics[width=16.8cm]{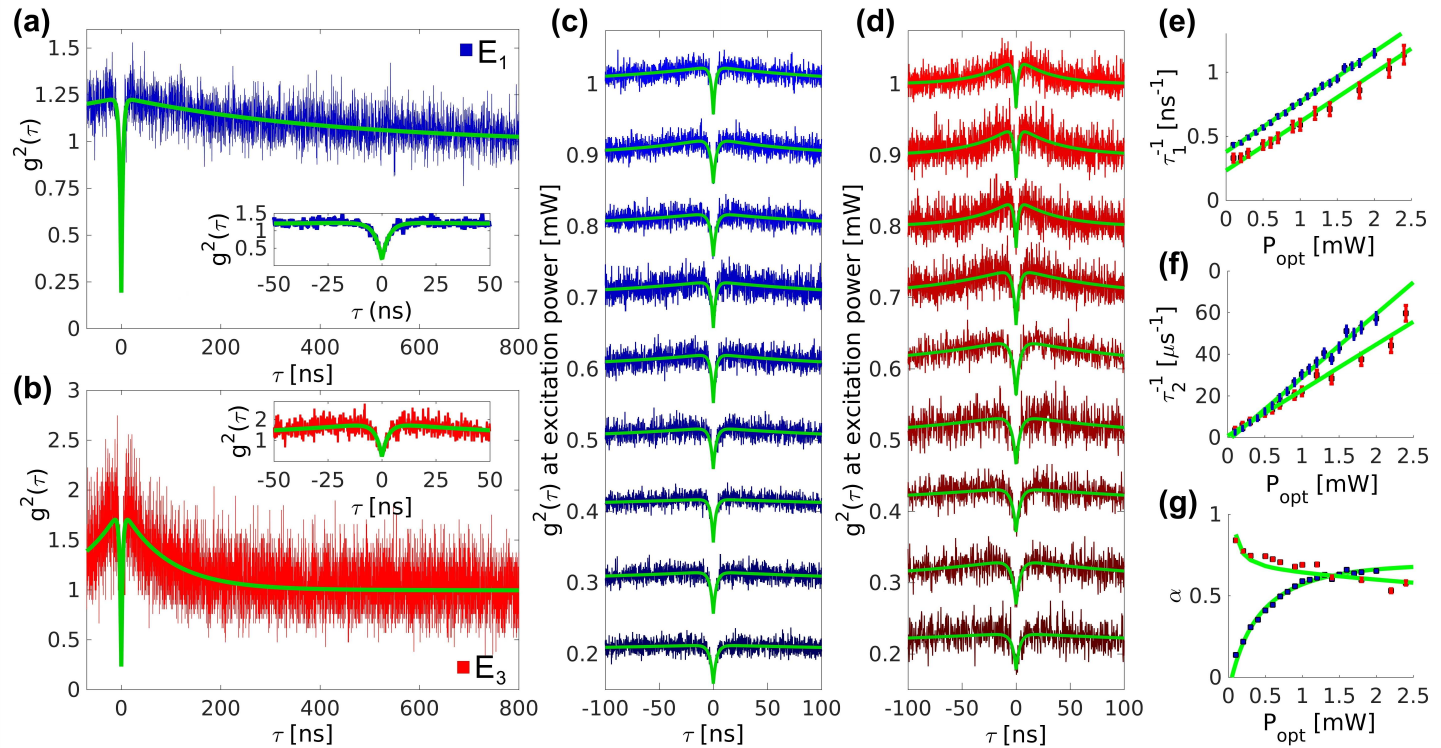}
	\caption{(a), (b) Second-order autocorrelation histograms of (a) E$_1$ and (b) E$_3$; (c), (d) power-dependent second-order autocorrelation histograms of (c) E$_1$ and (d) E$_3$; (e), (f) rreciprocal values of fitting parameters (e) $\tau_1$ and (f) $\tau_2$ for the second-order autocorrelation fit [Eq.~\ref{eq:g2}]. The power-dependent reciprocal values of $\tau_1$ and $\tau_2$ are subsequently linearly fitted (green) and reflect the decay rates of the excited and metastable states at the crossover of the extrapolated linear fit with the reciprocal time axis. (g) Power-dependent fitting parameter $\alpha$ accounting for the nonradiative transitions via the metastable state in comparison with the value calculated in Supplement 1 (green). Error bars in (e)--(g) represent the standard deviation calculated from the covariance matrix of each fitting routine.}
	\label{LevelAnti}
\end{figure*}

For a better understanding of internal electron dynamics and the level structure, we measured the photon statistics of the emitter with an HBT interferometer. The data presented here focus on the blueshifted emitter, E$_1$, and the redshifted emitter, E$_3$, as exemplary of the broad emitter distribution. Figs.~\ref{LevelAnti}(a) and \ref{LevelAnti}(b) show the normalized second-order autocorrelation histograms [g$\mathtt{^{(2)}}(\tau)$, Eq.~\ref{eq:g2}] without background correction, which confirm that the emission is dominated by a single emitter (g$\mathtt{^{(2)}}(0) < 0.5$). The insets show a magnified region near $\tau=0$. These histograms indicate a strong bunching for E$_3$, but only a weak bunching for E$_1$. These photon bunching features suggest the presence of a dark metastable state, as analyzed below. 

We use a rate equation analysis to model these autocorrelation measurements. A two-level model cannot capture the bunching behavior near $\tau=0$. A three-level model with pump-power dependent transition rates does capture the essential features of the second-order autocorrelation histograms, as seen in the green-curve fits in Fig.~\ref{LevelAnti}(a,b). This three-level model allows direct comparison with other well-studied emitters, such as the nitrogen-vacancy (NV) center in diamond~\cite{2006.PRB.Manson.NV_model} and the carbon antisite-vacancy pair in SiC~\cite{SiCRT}. 

The three-level system considered here consists of ground ($\ket{g}$), excited ($\ket{e}$), and metastable ($\ket{m}$) states. These states are coupled by the transition rates labeled in Fig.~\ref{Int}(a) and mathematically described by

\begin{equation}
\left(\begin{array}{c} \dot{p_g} \\ \dot{p_e} \\ \dot{p_m} \end{array}\right)
= \left(\begin{array}{ccc} 
-\gamma_{ge} 	& \gamma_{eg} 				& \gamma_{mg} 	\\
\gamma_{ge} 	& -\gamma_{eg}-\gamma_{em} 	& 0 			\\
0			 	& \gamma_{em} 				& -\gamma_{mg} 	\\ 
\end{array}\right)
\left(\begin{array}{c}
p_g \\ p_e \\ p_m
\end{array}\right).
\label{eq:rate}
\end{equation}

Eq.~\ref{eq:rate} describes a typical three-level system with the transition rates [$\gamma_{ij}$ between different levels $i,j \: \epsilon \: (g,e,m)$ with time-dependent populations $p_i$]~\cite{1983_RateE3L_Loudon_Oxford}. Electronic transitions in Fig.~\ref{Int}(a) account for contributions from excitation ($\gamma_{ge}$) and radiative decays ($\gamma_{eg}$) and nonradiative decays ($\gamma_{em}$, $\gamma_{mg}$). The transitions from the ground state to the metastable state ($\ket{g} \rightarrow \ket{m}$: $\gamma_{gm} \approx 0$) and from the metastable state to the excited state ($\ket{m} \rightarrow \ket{e}$: $\gamma_{me} \approx 0$) are assumed to be insignificant and are hence neglected.

Calculations in Supplement 1 of the decay ratio and radiative and nonradiative decays reveal that more than $90~$\% of all decays are radiative, allowing us to assume that the transition rate between $\ket{g}$ and $\ket{e}$ is significantly more dominant than the decay rate from $\ket{e}$ and $\ket{m}$ ($\gamma_{eg} \gg \gamma_{em}$). The $g^{(2)}(\tau)$ for a three-level system can thus be expressed in a first-order approximation as $g^{(2)}(\tau)=\lim_{t\rightarrow\infty} p_{e}(\tau)/p_e(t)$~\cite{G2For,2015_PhotoNV_Berthel_PRB, CrSat}. 

Evaluation of the latter approximation of $g^{(2)}(\tau)$ using Eq.~\ref{eq:rate} and the initial condition of the occupation of the ground state [$p_g(0)=1$] results in 
\begin{equation}
g^{2}(\tau) \approx 1-(1+\alpha) \exp(-\frac{\tau}{\tau_{1}}) + \alpha \exp(-\frac{\tau}{\tau_{2}}),
\label{eq:g2}
\end{equation}
with fitting parameters: $\tau_1(P_{\rm opt})$, $\tau_2(P_{\rm opt})$, and $\alpha(P_{\rm opt})$. These parameters result from the power-dependent decay rates from Eq.~\ref{eq:rate} , and thus depend on the respective excitation powers ($P_{\rm opt}$). The calculation of the individual decay rates and their power dependence is contained in Supplement 1.

Fitting parameter $\tau_1$ describes a two-level system considering transitions between the ground and excited state. Assuming a single excited state, $\tau_1$ can be linearly approximated as indicated in Fig.~\ref{LevelAnti}(e). $1/\tau_1$ shows a linear power dependence and can be considered as the lifetime of the excited state at zero optical excitation power ($P_{\rm opt}=0$)~\cite{G2For,CrSat}. The linear fit in Fig.~\ref{LevelAnti}(e) yields a lifetime for E$_1$ (E$_3$) of about 3.33$~\pm$0.2$~$ns (4.4$~\pm$0.3$~$ns).

The second fitting parameter, $\tau_2$, reflects the behavior of the metastable state. The fit in Fig.~\ref{LevelAnti}(f) reveals a lifetime, $\tau_{2}(0)$, of the metastable state of 675$~\pm$100$~$ns (900$~\pm$200$~$ns) for E$_1$ (E$_3$).

The third parameter of Eq.~\ref{eq:g2}, $\alpha$, plotted in Fig.~\ref{LevelAnti}(g) for E$_{1}$ and E$_{3}$, accounts for the nonradiative decays and transitions via the metastable state. $\alpha$ reveals the bunching of an emitter and thus is an indicator of the probability of intersystem crossing. Fig.~\ref{LevelAnti}(g) reveals that E$_1$ acts at low optical excitation power as a two-level system, while E$_3$ is strongly bunched characteristic for a three-level system with intersystem crossing.

Additional fluorescence lifetime measurements support the three-level model. Under pulsed excitation an excited state will decay exponentially [$I(t)\sim\exp{(-t/\tau)}$]. A single-exponential equation fits the decay rate of one single state with a time-dependent PL intensity term, $I(t)$, and with a fitting parameter $\tau$ that represents the lifetime of the spectrally detectable excited state. Fig.~\ref{Int}(b) shows the excited state lifetime measurement (wavelength, 532$~$nm; pulse length, 1$~$ns; frequency, 40$~$MHz). The optimized single-exponential fit indicates that there is only one excited state with a lifetime of 3.45$~\pm$0.1$~$ns for E$_1$ and 4.65$~\pm$0.3$~$ns for E$_3$. The measured lifetimes differ with the obtained fitting of $\tau_1$ and $\tau_2$ of the power-dependent g$^2$($\tau$), shown in Fig.~\ref{LevelAnti}(e), by only 0.2$~$ns, well within the error, supporting that a three-level system is a suitable model for the investigated emitters. 

\subsection{Photoluminescence Intensity}
\begin{figure}[ht!]
	\centering
	\includegraphics[width=8.5cm]{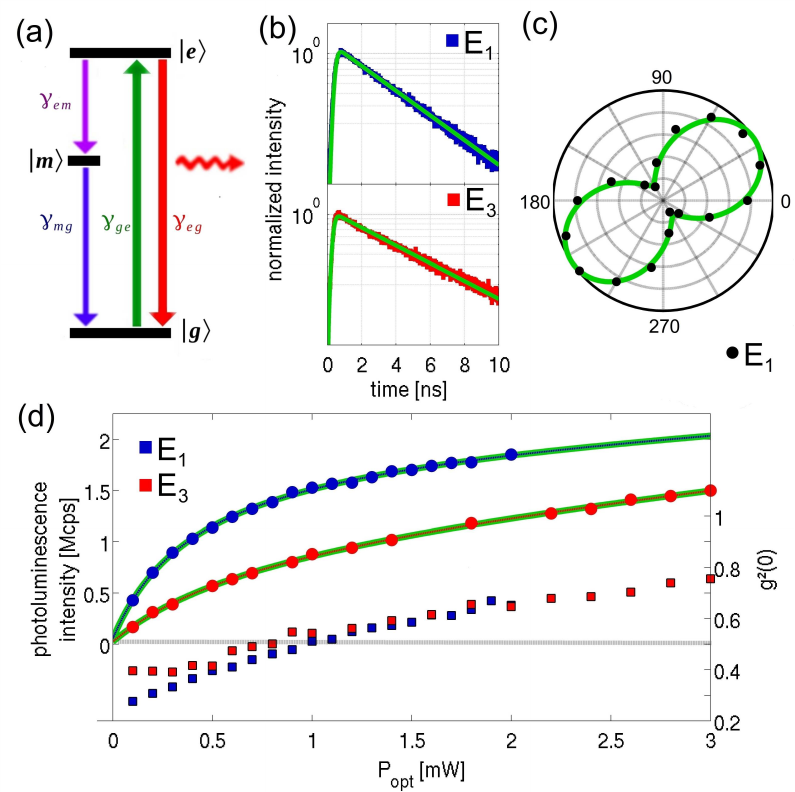}
	\caption{(a) Jablonski diagram of a three-level system with ground state ($\ket{g}$), excited state ($\ket{e}$), and metastable state ($\ket{m}$); (b) lifetime measurements fitted with a single exponential decay function; (c) polar plot of PL as a function of excitation laser polarization; (d) PL intensity measurements of the emitters evaluated at discrete excitation power to achieve maximal emission, fitted with Eq.~\ref{eq:sat} (green). Blue and red squares show the acquired background during PL measurement based on second-order autocorrelation histograms, with corresponding axis on the right-hand side. Both emitters show $g^2(0)<0.5$ up to an excitation power of $\sim$1$~$mW.}
	\label{Int}
\end{figure}

We studied the polarization absorption and emission properties by exciting the emitters with linearly polarized light. The excitation and emission polarization are individually measured and evaluated. The measurement method is further explained in Supplement 1. The emitters reveal parallel excitation and emission polarization. The collected data are fitted to $\sin^2(\theta+\phi)$, consisting of an angular parameter $\theta$, representing the rotation of the linear polarization of the pump laser, and $\phi$, the orientation of the emitter, relative to an arbitrary axis in the basal plane. Fig.~\ref{Int}(c) shows a polar plot of the excitation-polarization-dependent PL intensity (without background subtraction) of E$_1$. The histogram in Fig.~\ref{SpecCold}(b) summarizes these polarization measurements repeated for more than 20 emitters, which reveals only two linear polarization states, defined as $\phi_1=45^{\circ}$ and $\phi_2=135^{\circ}$.

A comparison between the obtained count rates with parallel and orthogonal linear laser excitation relative to the emitter polarization, and hence the minimum and maximum count rates in the polar plot in Fig.~\ref{Int}(c), reveals that all emitters are almost perfectly situated in a plane parallel to the sample surface, called the basal plane. More advanced measurements and calculations of the spatial orientation, elaborated in Supplement 1, support that the emitter's polarization is almost parallel to the basal plane, which is the orthogonal plane to the \textit{c} axis, indicated in Fig.~\ref{Figure1}(a). 

We next studied the maximum emission rates of E$_1$ and E$_3$ with optimized excitation polarization. Fig.~\ref{Int}(d) shows the saturation behavior of both emitters and the associated background based on second-order autocorrelation histograms. The $g^2(0)$-based background measurements reveal single-photon characteristics for both emitters up to excitation powers of $\sim$1$~$mW. To optimize the photon collection efficiency, these measurements were performed on a different confocal setup (Nikon, oil immersion, 1.3$~$NA, dichroic filter with cutoff wavelength of 552$~$nm, 560$~$nm longpass filter). The emission was equally distributed between two APDs to avoid saturation of the detectors.

For three-level systems, Eq.~\ref{eq:sat} below describes the power-dependent ($P_{\rm opt}$, measured after the objective) emission count rate ($R_{\rm COL}$)~\cite{2012.Cambridge.Novotny.QE, SatFor, RFor}, plotted in Fig.~\ref{Int}(d) and further described in Supplement 1. The equation is corrected with a linear power-dependent background slope ($a_{\rm BG}$) and a constant parameter for dark counts ($a_{\rm D}$):

\begin{equation}
R_{\rm COL}(P_{\rm opt}) =\frac{R_{\rm INF}P_{\rm opt}}{P_{\rm SAT}+P_{\rm opt}}+a_{\rm BG} P_{\rm opt}+a_{\rm D}.
\label{eq:sat}
\end{equation}

As seen in Fig.~\ref{Int}(d), the emitters exhibit high brightness, with more than one million cps detected on each APD. The calculated saturation counts ($R_{\rm INF}$) and corresponding saturation power ($P_{\rm SAT}$) of E$_1$ are estimated to be at 1.942$\times$10$^6~$cps and 0.425$~$mW, and they are 1.154$\times$10$^6~$cps and 0.78$~$mW for E$_3$. This is one of the brightest emitters in any bulk material reported to date, and among the brightest SPEs even if compared to color centers in diamond nanocrystals~\cite{CrSatN, RFor}. The PL intensity at saturation of an NV color center in bulk diamond on the same setup under the same conditions revealed that the emitters hosted by bulk SiC are about one order of magnitude brighter. Therefore, the emitters are comparable to a NV center enhanced via an optical nanostructure~\cite{OR1}.

\subsection{Photoluminescence at Cryogenic Temperatures}
We performed low-temperature (18$~$K sample temperature) PL measurements in a closed-cycle cryostat (Janis) to study phonon coupling characteristics.

\begin{figure}[ht!]
	\centering
	\includegraphics[width=8.5cm]{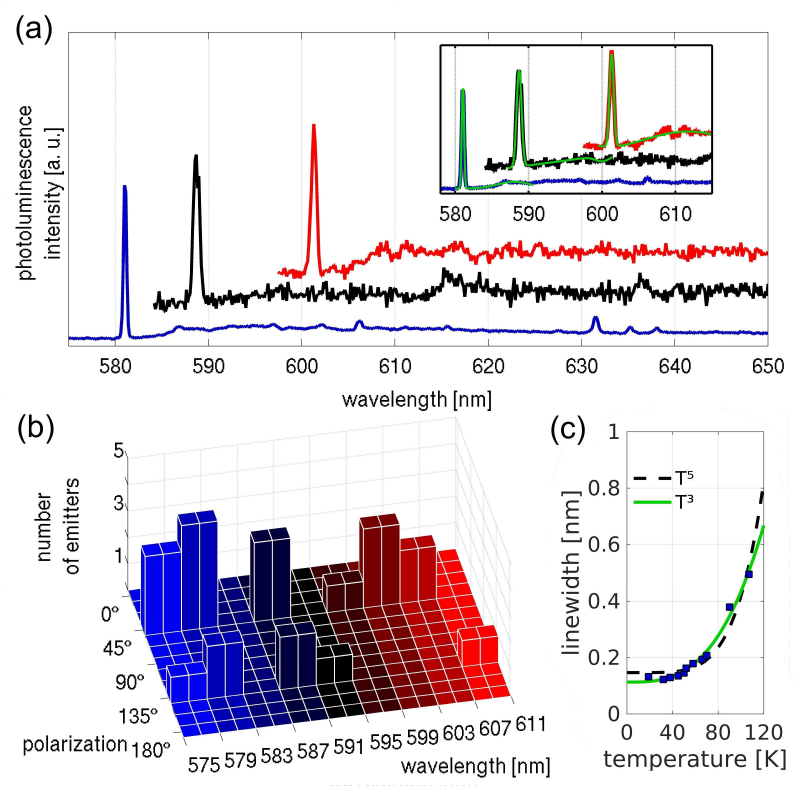}
	\caption{(a) PL spectrum of E$_1$ (blue), E$_2$ (black), and E$_3$ (red) in an environment at 18$~$K. Inset: magnified spectral range containing ZPL (E$_1$: 581.2$~$nm, E$_2$:  588.9$~$nm, E$_3$: 601.5$~$nm) and first phonon modes, highlighted with green Gaussian fits. (b) Histogram of ZPLs with the corresponding polarization of various room-temperature emitters at different wavelengths recorded at cryogenic temperatures. (c) Plot of the linewidth broadening with increasing temperature (T) of the dominant peak at 581.4$~$nm (E$_1$, blue squares). A free fit ($\propto T^x$) yields $x=3.02$ indicating a $T^3$ dependence (solid green line). For comparison the dashed black line shows the case for T$^5$.}
	\label{SpecCold}
\end{figure}

Fig.~\ref{SpecCold}(a) shows low-temperature PL spectra of the three emitters introduced in Fig.~\ref{Figure1}(c). Common to all low-temperature spectra is a strong peak, which we attribute to the ZPL, and a red shifted ($\sim$ 8$~$nm) phonon broadened peak, indicated in the magnified cutout in Fig.~\ref{SpecCold}(a). The ZPL emission wavelengths are illustrated in Fig.~\ref{SpecCold}(b) as a histogram and reveal a spread from $\sim 575$ to $610~$nm. There is no recognizable dependence between ZPL and polarization state.

Temperature-dependent ZPL linewidth broadening reveals information about the dephasing mechanisms of the emitter, as well as the influence of the host crystal on the defects. Fig.~\ref{SpecCold}(c) shows the temperature dependence of the emitter linewidth of the most dominant peak of E$_1$ at 581.4$~$nm fit with Gaussian functions (spectrometer resolution: $0.07~\pm0.01~$nm). At 18$~$K we determine a linewidth of 0.11$~\pm$0.01$~$nm. The lines were stable over the measurement period, and no diffusion or blinking on the millisecond time scale was observed. The linewidth broadening with increasing temperature is very well approximated with a $T^3$ behavior, shown in Fig.~\ref{SpecCold}(c), similar to the silicon vacancy and other defects in diamond~\cite{2013.NJP.Neu.SiVLT}. The $T^3$ dependence is caused by field fluctuations due to phonon-induced dislocations of crystal defects and color centers~\cite{1999_AIP_Hizhnyakov_Dephasing}, and is often reported for emitters with high inhomogeneous broadening~\cite{2013.NJP.Neu.SiVLT}. This is significantly different from the $T^5$ dependence of the NV$^-$ in diamond, which originates from the dynamic Jahn-Teller effect~\cite{2002_PSS_Hizhnyakov_ZPLJahnTeller, DJT}. 

Our results do not show spectral diffusion of the reported emitters on the millisecond time scale. However, spectral diffusion dynamics might have time scales on the nanosecond or microsecond scale~\cite{PhysRevLett.110.027401}, causing inhomogeneous broadening of the ZPL. It has been shown for other solid-state SPEs that spectral diffusion can be reduced significantly if the SPE environment is of high quality or prepared in a certain way~\cite{2014_NL_Lunkin_Dephasing}, indicating that such processes for the presented emitter could be controlled in future experiments. As has been shown with the SiV and NV centers in diamond, resonant excitation coupled with carefully prepared samples reduces spectral diffusion and pure dephasing, enabling high-visibility quantum interference~\cite{2014.NJP.Jelezko.SiVDiff}. We are hopeful that similar principles will apply to this emitter.

The population of the phonon sideband (PSB), expressed in Eq.~\ref{eq:DW} by the Debye-Waller factor (DWF), can be estimated with the ratio of the ZPL PL emission, $I_{\rm ZPL}$, relative to the total PL emission, $I_{\rm TOT}=I_{\rm ZPL}+I_{\rm PSB}$, which is the combination of the ZPL PL emission and the phonon-broadened PL, $I_{\rm PSB}$. The DWF is calculated by separately fitting the ZPL and PSB peaks with Gaussian fit functions:

\begin{equation}
\text{DWF} = \frac{{I_{\rm ZPL}}}{{I_{\rm TOT}}}.
\label{eq:DW}
\end{equation}

At cryogenic temperatures we determine for E$_1$ a DWF of 33$~\pm 1~$\%. NV color centers yield a DWF of only about $3~$\%~\cite{2006.PRB.Manson.NV_model} but the silicon-vacancy center in diamonds has a DWF of more than $70~$\%~\cite{2011.NJP.Zaitsev.SiVDWF}. A higher DWF is preferred for achieving better single-photon emission characteristics~\cite{2008.MatTod.Prawer.DWF}, and could be directly enhanced via coupling to an optical cavity~\cite{DStruc}.

\section{Discussion}
Finally, we discuss the origin of these emitters. We attribute the investigated SPEs to the same emitter type due to similar low- and room-temperature PL spectra, the linear polarization data, and the matching of a three-level system as the underlying model. We denote the emitter type as intrinsic due to the high density of emitters in high-purity SiC. Furthermore, the SPEs are observed in both electron-irradiated and unirradiated annealed samples, with a higher density after electron irradiation.

To find a possible origin of the emitter, three major indicators are considered. (1) The investigated SPE's spectral components are in the red visible range. (2) To create photostable SPEs, it is necessary to anneal the SiC samples in an inert environment of at least 600$^{\circ}~$C. There is no observation of degradation regarding photostability or changes in emitter density up to annealing temperatures of at least 900$^{\circ}~$C. (3) There are two linear, orthogonal polarization states.

In the red visible spectral range, two SPEs in 4H-SiC are known and have been characterized: the carbon antisite-vacancy pair, characterized by Castelletto \textit{et al.}~\cite{SiCRT}, and a thus far unknown emitter type reported by Lohrmann \textit{et al.}~\cite{2015_NatCom_Lohrmann_SiC6H}, presumably hosted by 6H-SiC inclusions. The carbon antisite-vacancy pair differs from the measured SPE, especially by a redshifted first ZPL (648.5$~$nm), a shorter excited state lifetime, and a stronger irradiation-density dependence. In contrast to the described SPE, the latter, yet unidentified emitter type exists in three different linear polarization states ($\phi_1=30^{\circ}$, $\phi_2=90^{\circ}$, $\phi_3=150^{\circ}$) instead of two. Furthermore, its ZPLs are distributed over a broader spectrum ($\sim 550 - 750~$nm) and the saturated PL intensity is significantly smaller.

We exclude, as possible underlying defect structures, single carbon (C) and silicon (Si) vacancies, since C vacancies tend to anneal out at temperatures above 500$^{\circ}~$C~\cite{CVD} and Si vacancies~\cite{SiVSiC1, SiVSiC2, SiVSiC4} emit in the near IR. The D$_{I}$-defect~\cite{D1}, D$_{II}$-defect~\cite{D2}, and di-vacancies~\cite{DiV1} show optical signatures in different spectral ranges, which eliminates them as possible emitter origins. Antisites, which are Si (C) atoms replaced by C (Si) atoms, show very low formation energies, and hence they are the most common defects in SiC~\cite{ANT}. Nevertheless, annealing temperatures around $200^{\circ}~$C to $300^{\circ}~$C cause antisites to become mobile, and they eventually recombine. In conclusion, none of the optically characterized emitters or described defect structures in 4H-SiC seem to match the observed and measured features of the emitter.

Based on the conflict between the possible polarization states in hexagonal polytype hosts [$60^\circ$ separation between states, apparent in Fig.~\ref{Figure1}(a)] and the two observed orthogonal polarization states of the emitter, we propose a polytype inclusion. In the case of 3C-SiC, the most common cubic polymorph in SiC, just a single plane needs to be skipped and the violation in terms of the polarization is resolved because of the orthogonality between different bonds~\cite{2003_APL_Bai_SiCPC, 2015_NatCom_Lohrmann_SiC6H}, as further discussed in Supplement 1. During the production of SiC wafers by Cree, Inc., epitaxy defects such as 3C inclusions can be incorporated, particularly where the step flow was interrupted during epitaxy layer growth~\cite{CREE_SiCWafer}.

\section{Conclusion}
We described an exceptionally bright visible-spectrum SPE in silicon carbide. The emitters exhibit count rates up to 2$\times$10$^6~$cps at saturation at room temperature---making them one of the brightest sources of single photons from unpatterned bulk material. The emitters are abundant in a nonirradiated, annealed high-purity 4H-SiC sample. Furthermore, their linear polarization, narrow linewidth, and stability under constant laser excitation are promising attributes for applications in integrated nanophotonic technologies.

\section*{Acknowledgments}
Sinan Karaveli, Edward H. Chen, and
Michael Walsh are acknowledged for helpful discussions, and Lei Zhu and Sinan Karaveli for revising the paper.\\
I. A. is the recipient of an Australian Research Council Discovery Early Career Research Award (Project Number: DE130100592).\\
This work was supported in part by the Army Research Laboratory CDQI program and the Air Force Office of Scientific Research Multidisciplinary University Research Initiative (FA9550-14-1-0052) and the Asian Office of Aerospace Research and Development (AOARD) (FA2386-15-1-4044).

\bibliography{Citation}
\end{document}